# Anisotropic Subdiffractive Solitons

Ramon Herrero[1], I. V. Barashenkov[2], N. V. Alexeeva[2], and Kestutis Staliunas[1,3],

[1]*Departament de Física i Enginyeria Nuclear,Universitat Politècnica de Catalunya, Colom 11, 08222 Terrassa, Spain*
[2]*Department of Mathematics, University of Cape Town, Rondebosch 7701, South Africa*
[3]*Institució Catalana de Recerca i Estudis Avançats (ICREA),*

**Abstract**

We study solitons in the two-dimensional defocusing nonlinear Schrödinger equation with the spatio-temporal modulation of the external potential. The spatial modulation is due to a square lattice; the resulting macroscopic diffraction is rotationally symmetric in the long-wavelength limit but becomes anisotropic for shorter wavelengths. Anisotropic solitons -- solitons with the square *(x,y)*-geometry -- are obtained both in the original nonlinear Schroedinger model and in its averaged amplitude equation.

PACS numbers: 03.75.Kk, 05.45.Yv

## 1. Introduction

The Nonlinear Schrödinger (NLS) equation without an external potential does not support bright solitons in the defocusing (repulsive) case, where the cubic nonlinearity and coefficient of the diffraction term have opposite signs. A spatially periodic modulation of a stationary external potential can invert the sign of the diffraction, which allows the formation of localized structures, the so-called gap solitons [1]. The spatio-temporal modulation of the potentials gives an additional degree of freedom in the manipulation of diffraction. For example, the "shaken" [2] and the "blinking" [3] potentials can go so far as to tune the diffraction coefficient exactly to zero. The second spatial-derivative terms are absent then, and the subdiffraction, i.e. diffraction of a higher order, becomes responsible for the spreading of light beams or wave packets.

The subdiffraction of wave packets can be balanced by nonlinearity, leading to the formation of the so-called subdiffractive solitons. The subdiffractive solitons have been predicted for the "shaken" [4] and "blinking" [3] lattices in the one space-dimensional (1D) case. More recently, the flattening of diffraction surfaces, and formation of subdiffractive solitons, was proposed for the 2D and 3D NLS equations, with the 2D and 3D spatial modulation of the potential [5].

The spatially modulated NLS equations are no longer isotropic (no longer rotationally symmetric). The solitons resulting from the balance between nonlinearity and anisotropic diffraction, inherit this low spatial symmetry. The aim of the present paper is to study the anisotropic solitons in the 2D nonlinear Schrödinger equation of defocusing type, with the square-symmetric spatial potential subject to a harmonic temporal oscillation. This type of modulation has already been used in the study of solitons and vortices in the two-dimensional and quasi one-dimensional lattices [6]. Unlike Ref.[6], we consider the system to be close to the subdiffractive regime and with the set of normalized parameters ensuring the maximum degree of anisotropy.

The outline of the paper is as follows. In the next section we report results of the numerical simulation of the original NLS with a potential periodically modulated in space and time. Here the small-scale space and time variations of the potential appear explicitly. We demonstrate the existence of stable solitons as envelopes of the Bloch waves in this microscopic model. In section 3, we introduce the macroscopic equation governing the envelope of the Bloch mode. Here, the dependence on small space and time scales is averaged out, so that the field changes on large scales only. In section 4 we employ a variational method to show that the macroscopic equation supports anisotropic solitons. These predictions of the variational approach are confirmed in direct numerical simulations of the time-dependent macroscopic equation (section 5). Stable solitons are shown to exist when the propagation constant lies above a certain critical value. Finally, results of this study are summarized in section 6.

## 2. The microscopic model

The model we start with is the nonlinear Schrödinger (NLS) equation with an external potential periodic in space and time:

$$\frac{\partial \psi}{\partial t} = i\left[\nabla^2 + 4f \cos(\Omega_0 t) V_s(\vec{r}) - |\psi|^2\right]\psi, \quad (1.a)$$

$$V_s(\vec{r}) = \sum_{j=1}^{s} \cos(\vec{q}_j \cdot \vec{r}), \quad (1.b)$$

Here the potential $V_s(\vec{r})$ is formed by the superposition of $s$ one-dimensional cosinusoidal lattices with wave-vectors $\{\vec{q}_j\}_{j=1}^{s}$ (satisfying $|\vec{q}_j|=1$). In addition, it performs harmonic oscillation in time with the frequency $\Omega_0$. We focus on the case $s=2$, i.e. on the square lattice.

Physically, the model (1) describes the paraxial propagation of monochromatic light through a Kerr-nonlinear media with periodically and harmonically modulated refractive index – the so-called photonic crystals [7,8]. Eq.(1) corresponds to the defocusing nonlinear medium. Here $t$ denotes the longitudinal spatial coordinate, so that the system is 3-dimensional and time-independent. The same equation models the dynamics of the Bose-Einstein condensate (BEC) in a dynamical potential formed by square lattices "blinking" in time [3,5]. In the latter case the system is 2-dimensional in space and evolving in time. For the uniformity of the presentation, we adopt the "time" interpretation of the variable $t$ –- keeping in mind that in optical systems it actually stands for the longitudinal coordinate.

Both in the optical and atomic systems the modulation of the external potential on a short spatial scale produces a modification of the net diffraction of wave packets on a longer scale. This results either in the change of the sign of macroscopic diffraction [1], or in the elimination of the diffraction to the leading order. Here by the macroscopic diffraction we mean the net diffraction of the envelope of the Bloch modes, and the leading order is given by the first power of the Laplacian, $\nabla^2 = \partial^2/\partial X^2 + \partial^2/\partial Y^2$ [3,5]. The capitals $X$ and $Y$ denote large-scale space coordinates associated with the variation of the envelope of the Bloch modes.

For localized structures to be formed, the negative diffraction needs to be balanced by the nonlinearity corresponding to defocusing nonlinear materials in the optical case and repulsive bosons in the case of the Bose condensates. When the diffraction vanishes to the leading order, the next order of the diffraction, or subdiffraction, takes over in the corresponding macroscopic amplitude equations. The subdiffraction is described by the next higher powers of the second derivatives: $\partial^4/\partial X^4$, $\partial^4/\partial Y^4$ and $\partial^4/\partial X^2\partial Y^2$.

Solutions arising in such diffraction-manipulated nonlinear Schrödinger systems were scrutinized in the one dimensional case ($s=1$) [3]. The situation in 2D is more involved as the diffraction, managed by the lattices, can become anisotropic and introduce transverse optical effects such as the tunneling inhibition [9]. The higher rotational symmetry the lattice has (the larger the $s$), the higher order isotropy the diffraction surface $\omega(\vec{k})$ possesses in the wavenumber space (in the far field). For instance, the octagonal ($s$=4) lattice considered in [5] gives rise to fairly isotropic solitons.

However the modulations that are most relevant experimentally are due to the square symmetric lattices ($s$=2). Both in the BECs and in nonlinear optics, the square lattices are the easiest to implement. In this case the isotropy is granted only to the leading order in the wavenumber $\vec{k}$ [5]: the second spatial derivatives $\partial^2/\partial X^2$ and $\partial^2/\partial Y^2$ enter the NLS equation isotropically, whereas the forth derivatives $\partial^4/\partial X^4$, $\partial^4/\partial Y^4$ and $\partial^4/\partial X^2\partial Y^2$ enter in a non-isotropic way.

We have performed numerical simulations of the microscopic model (1) with the square-lattice potential. The split-step method was implemented in a $30\pi \times 30\pi$ box, discretized by a grid of 256×256 points. Solitons were seeded by a gaussian initial condition. The radiation produced while the solution was settling to a stationary state, was drained away by using the "absorbing" boundary conditions. These were emulated by adding a dissipative term with the coefficient gradually increasing towards the boundaries of the integration domain. Once the radiation transients were over, the damping term would be turned off and the simulations continued under periodic boundary conditions.

Stable solitons of equation (1) were found in a range of parameters $\Omega_0$, $f$ and $q_i$. An example is shown in Fig. 1. The soliton is obviously anisotropic, that is, does not exhibit radial symmetry. The anisotropy is obvious not only in the instantaneous snapshots of the $\psi$ field (Fig.1.a) but also in the time-averaged (Fig.1.c) and space-filtered (Fig.1.d) profiles, where the small-scale modulation has been eliminated. Both the time averaging and momentum filtering leave only the large space-scale modulation of the soliton intensity distribution, $|\psi(x,y)|^2$ (Fig.1.c,d).

The fact that the $\psi$ field (the microscopic field) has the square symmetry at the short scale imposed by the period of the potential is natural: the solution is symmetric because the coefficients of the equation (1) are symmetric. The square symmetry of the *envelopes* of the $\psi$ field -- that is, the symmetry on a scale longer than the period of the lattice -- is less trivial.

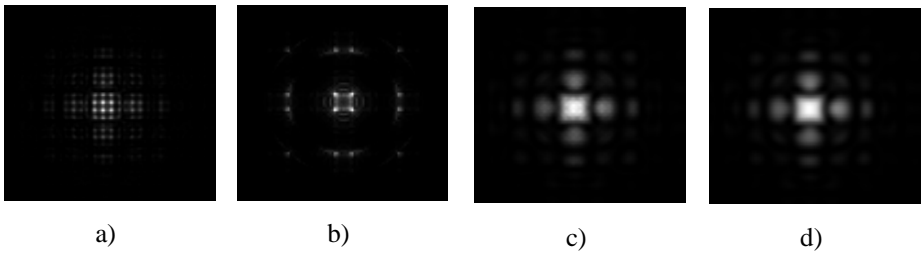

a)  b)  c)  d)

**FIG. 1.** The grayscale gradient map showing spatial intensity distribution of the field in the soliton, as obtained by numerical integration of the microscopic model (1). (a) The instantaneous intensity distribution $|\psi|^2$ on the *(x,y)* plane; (b) the intensity in the space of wavenumbers (spectral energy density); (c) the intensity on the *(x,y)* plane after averaging over one temporal period of the lattice oscillation; (d) the intensity on the *(x,y)*

plane with the side-band spectral components (components with $|k|>1/2$) filtered out. The parameters are: $f=0.028$, $\Omega_0=0.62$; the integration domain $(-15\pi,15\pi)\times(-15\pi,15\pi)$, and the wavenumber window in (b) is (-1.75,1.75) × (-1.75,1.75).

## 3. The macroscopic model

The amplitude equation for the envelope of the Bloch mode has been derived in [5]. The solution is sought in the form $\psi(\vec{r},t)=A(\vec{r},t)\psi_B(\vec{r},t)$, where $A(\vec{r},t)$ is the envelope of the spatio-temporal Bloch mode $\psi_B(\vec{r},t)$ of equation (1). The derivation assumes that only one Bloch mode is enveloped by the soliton; this is a legitimate assumption if the nonlinearity is weak. It also assumes that $f \ll 1$ (meaning a weak modulation of the potential) and $|1-\Omega_0| \ll 1$ (implying the proximity to the multiple cross-point of the diffraction paraboloids). More specifically, the following scaling laws were used: $1-\Omega_0=O(\delta)$, $4sf^2=O(\delta^3)$, where $0<\delta \ll 1$ is a small parameter [5].

In the square-lattice case ($s=2$) the amplitude equation reads

$$\frac{\partial A}{\partial t}=i\left[d_2\nabla^2+d_4\nabla_4-|A|^2\right]A \cdot \quad (2)$$

Note that the form $\nabla_4=\left(\partial^4/\partial X^4+\partial^4/\partial Y^4\right)$ appearing in (2) is not equivalent to the spatially isotropic operator $\nabla^4$. Unlike $\nabla^4$, this form does not possess any rotational symmetry. The diffraction coefficients are given by

$$d_2=1+8f^2(\Omega_0-1)^{-3}-64f^4(\Omega_0-1)^{-5}+\cdots \quad (3.a)$$

$$d_4=-32f^2(\Omega_0-1)^{-5}+\cdots \quad (3.b)$$

The linear dispersion relation of eq.(2) is $\omega(\vec{k})=-d_2(k_x^2+k_y^2)+d_4(k_x^4+k_y^4)$. The corresponding dispersion surfaces $\omega(\vec{k})$ are depicted in Fig.2. For the sake of comparison, we also show the corresponding dispersion surfaces for the microscopic model (1). As $f$ is increased, positions of the minima of $\omega(\vec{k})$ in the two models start to differ, but the general shapes of the two surfaces remain the same. The broad plateau around the origin in Fig.2 (a) and (c) indicates the vanishing of the low-order diffraction when $d_2=0$. On the other hand, a pronounced asymmetry of the surface in Fig.2(b) and (d) corresponds to positive $d_2$ values.

Rescaling the variables $X, Y$ and $A$ we can always arrange that $d_4=1$. The coefficient $d_2$ can be set to 1, 0 or -1 – depending on whether we are considering the regimes of normal diffraction, subdiffraction and antidiffraction. The anisotropy is most pronounced in the normal diffraction case ($d_2=1$); thus, the rest of the paper is devoted to this case.

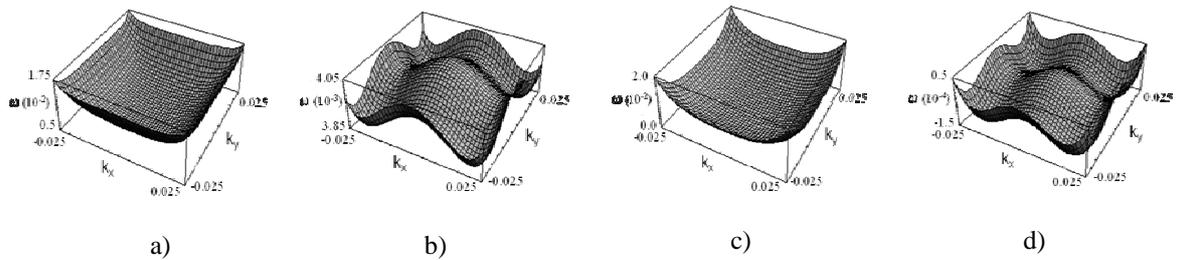

a)   b)   c)   d)

**FIG. 2.** Surfaces of the propagation constant of the Bloch mode, as obtained by the harmonic expansion of eq.(1) with a potential periodic in space and time. (a) $\Omega_0=0.95$ (the zero-diffraction case); (b) $\Omega_0=0.9$ (the normal diffraction regime). For comparison, the corresponding dispersion surfaces in the harmonic expansion for eq.(2) are plotted in (c) and (d). In panels (a) and (c), the potential has one broad minimum at the origin, while in (b) and (d), there are 4 minima and 4 saddles appearing along a square-shaped potential valley. In this figure, $f=0.01$.

## 4. Macroscopic solitons: variational approximation

Having demonstrated the existence of stable square-symmetric solitons in the NLS equation (1) and having established the relation between the microscopic model (1) and its macroscopic counterpart (2), we proceed to the analysis of the corresponding soliton solutions of equation (2).

We start with a variational approach where the soliton is approximated by the function

$$A(\vec{r})=A_0\exp\{-i\mu\tau-(x^2+y^2)/r_0^2\}\cos(kx)\cos(ky). \quad (5)$$

Here $\mu$ is the propagation constant and $A_0$, $r_0$, $k$ parameters of the soliton to be determined by the minimization of the Hamiltonian. In addition to the variational parameters used in the studies of the isotropic solitons (the width $r_0$ and the amplitude $A_0$), we have included the wavenumber of the macroscopic scale modulation, $k$. The purpose of the introduction of this additional degree of freedom is to emulate the anisotropy of the soliton.

The Hamiltonian for the equation (2) has the form

$$H = \int \left( d_2 |\nabla A|^2 - |\partial_{xx}^2 A|^2 - |\partial_{yy}^2 A|^2 + \frac{|A|^4}{2} - \mu |A|^2 \right) d\vec{r} \,. \tag{6}$$

Substituting the Ansatz (5) in the functional (6) and minimizing $H$ with respect to $A_0$, $r_0$, $k$, we obtain three algebraic equations. The equilibrium soliton parameters are given by roots of these equations; these can be expressed as functions of μ, with $\mu > 0.5$ (Fig.3).

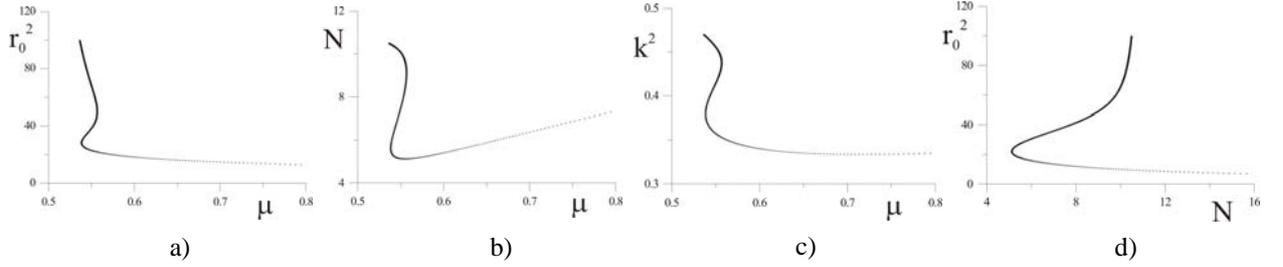

a) b) c) d)

**FIG. 3.** Roots of the variational equations. Diagrams show the width (a), energy (number of particles) (b) and wavenumber (c) as functions of the propagation constant $\mu$. In (d), the width is plotted as a function of the number of particles.

When $\mu$ is large and $k \neq 0$, the variational equations have only one root corresponding to a soliton with a large amplitude and small width [Fig 3 (a) and (d)]. In the interval $0.538 < \mu < 0.557$, there are three coexisting roots. As $\mu$ drops down to 0.5, the amplitude decreases, width grows and the soliton acquires a checkered profile.

## 5. Numerical simulations of the macroscopic model

In order to verify the conclusions of the variational analysis, we integrated equation (2) numerically. Here we utilized the same split-step approach as for eq. (1). This time our grid included 128×128 points; this provided a sufficient resolution in the absence of the short-wavelength spatial scale. Anisotropic solitons were found in each of the three possible regimes: normal diffraction, subdiffraction and antidiffraction. In what follows, we focus on the normal diffraction ($d_2 = 1$) as the most anisotropic case.

The energy of the resulting soliton solutions was reduced, in steps, by adding a homogeneous damping term and turning it off once the energy has dropped by the desired amount. We also imposed absorbing boundary conditions on the *(X,Y)* plane. Here the energy of the soliton (referred to as the number of particles in the BEC context) is defined by $N=\int |A|^2 dXdY$.

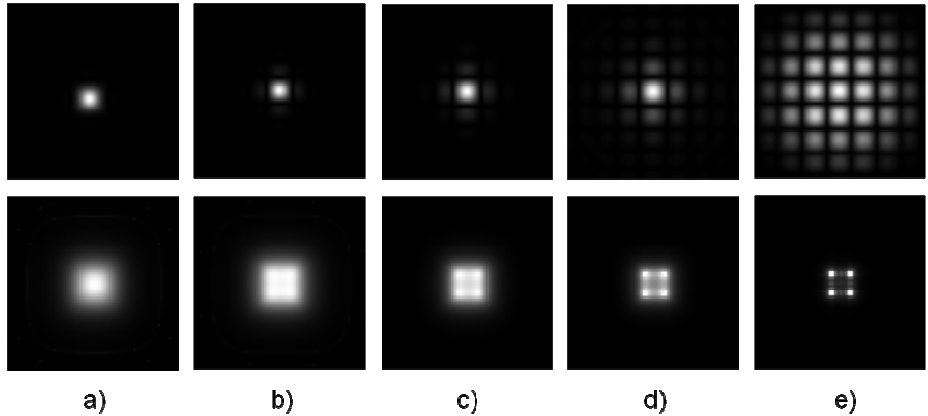

a) b) c) d) e)

**FIG. 4.** The grayscale colourmap showing the energy distribution in the soliton, as obtained by the numerical integration of the macroscopic model (2). The upper row depicts the distribution of $|A|^2$ on the *(X,Y)*-plane while the bottom row shows the corresponding energy distribution in the wavenumber space (the spectral energy density). The total energy (the total number of particles captured in the soliton) decreases from left to right: (a) N=37; (b) N=18; (c) N=10,; (d) N=6.15; (e) N=5. Parameters of the macroscopic equation (2): $d_2 = d_4 = 1$.

Fig.4 shows the grayscale gradient maps of solitons with different energies on the (*X,Y*)-plane (upper row) and on the plane of wavenumbers (bottom row). Solitons with high energies (large *N*) have large amplitudes and small widths, with little modulation; they appear almost isotropic. As the soliton energy is reduced, its amplitude decreases and width grows. The localized structure adds an increasing number of spatial undulations and develops strong asymmetry, acquiring a checkered shape (Fig.4e). (Note, in particular, the panel (d) picturing a low-energy soliton with parameters close to those of the microscopic distributions in Fig.1.)

The soliton's anisotropy properties admit a simple interpretation in terms of the dispersion surface $\omega = \omega(\bar{k})$. Indeed, since solitons with high energies *N* have narrow widths, they have a broad spatial spectrum, without any preferred wave vector. This implies a high degree of isotropy. On the other hand, low-energy solitons have a smaller µ (µ>0), with $-\mu$ lying just below the four minima of the surface $\omega = \omega(\bar{k})$. Consequently, as the energy of the soliton decreases, it becomes dominated by four wave vectors pertaining to the minima of the surface and therefore, extremely anisotropic.

Surprisingly, our simulations did not find stable solitons in the limit of small propagation constants. As $\mu$ is decreased down to 0.5, the soliton's width blows up while its amplitude decays to zero (Fig.5). The stable small-amplitude soliton ceases to exist. The nonexistence of stable subdiffractive solitons with small propagation constants contrasts the case of the "standard" NLS equation with the second-order derivatives and self-focusing (attractive) cubic nonlinearity. The second-order (isotropic) self-focusing NLS has stable bright solitons with arbitrary negative propagation constants, from zero to minus infinity.

We should also mention that no bistabilities have been detected during our energy-reduction process; this indicates that all observed solitons correspomd to the same branch in Fig.3. Solitons corresponding to other stationary branches have not been observed numerically.

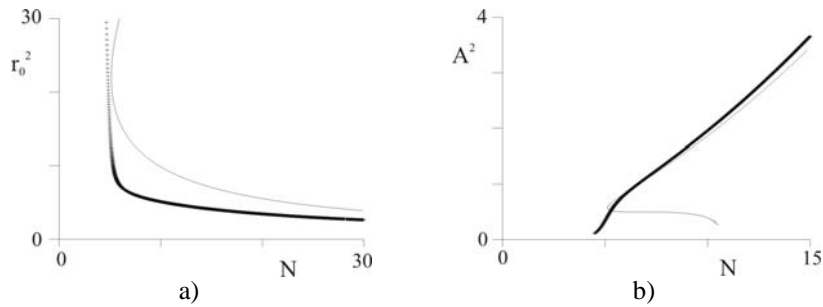

**FIG. 5.** The width (a) and the amplitude (b) of solitons as a function of energy. Solid line and dots: results of the numerical integration of the macroscopic evolution equation (2). Thin lines: variational approximation.

Fig.5 compares the amplitude and width of the soliton obtained via the numerical integration of the macroscopic equation (2) to the variational results. The numerical and variational results for the soliton amplitudes are found to be in excellent agreement. On the other hand, the numerical widths are always below their variational counterparts. One reason for this discrepancy is the finiteness of the spatial simulation domain. Another one is the long time the solution takes to settle to the stationary state as $\mu$ approaches the stable soliton existence threshold of 0.5.

## 6. Conclusions

We have shown that stable anisotropic solitons exist in the Kerr-nonlinear systems with the spatiotemporal modulation of the external potential. The potential has a form of a lattice with the lowest possible symmetry, i.e. a square lattice. We have established the existence of solitons both within a micro- and macroscopic model. The microscopic solitons are obtained in direct numerical simulations of the nonlinear Schroedinger equation. The macroscopic solitons are obtained variationally, via the Hamiltonian minimization; the variational results are verified in direct numerical simulation of the macroscopic evolution equation. Solitons are stable and almost isotropic in the limit of high energies (large numbers of particles). Solitons with low energies have large widths and checkered modulated shapes. Stable checkered solitons cease to exist as their energies and propagation constants decrease below certain critical minima.

**Acknowledgments**

This study was supported by the acciones integradas HS2007-0010. R.H. and K.S. were supported by the Spanish Ministerio de Educación y Ciencia and European FEDER (grant FIS2008-06024-C03-02). I.B. and N.A. were supported by the NRF of South Africa (grants UID 65498, 68536 and 73608).